\documentclass[aps,noshowpacs,nofootinbib,onecolumn,eqsecnum,floatfix,amsmath,amssymb]{revtex4}
\usepackage{amsbsy}
\usepackage{amsfonts}
\usepackage{amssymb}
\usepackage{amsmath}
\usepackage{bm,latexsym}
\usepackage{mathrsfs}
\usepackage{graphicx}
\usepackage{epsfig}

\def\Journal#1#2#3#4{{#1} {\bf #2}, #3 (#4)}
\def\AA{{Astron. Astrophys.}}
\def\AJP{{Am. J. Phys.}}

\def\APJ{{Astrophys. J.}}

\def\ARAA{{Annu. Rev. Astrophys.}}
\def\EJP{{Eur. J. Phys.}}

\def\CQG{{Class. Quant. Grav.}}

\def\GRG{{Gen. Relativ. Gravit.}}

\def\PRL{Phys. Rev. Lett.}

\def\PRD{{Phys. Rev.} D}
\def\PR{{Phys. Rep.}}

\def\ba{\begin{eqnarray}}
\def\ea{\end{eqnarray}}
\def\be{\begin{equation}}
\def\ee{\end{equation}}

\def\={\mathrel{\widehat\mathalpha{=}}}
\def\punto#1{\rlap{\raise.5ex\hbox{\char'27}}{#1}}

\newcommand{\cero}[1]{\stackrel{_o}{#1}}

\begin{document}





\title{The gravitational light shift and the Sachs-Wolfe effect}


\author{Cesar Merl\'\i n}
\email{cesar.merlin@nucleares.unam.mx}
\affiliation{Instituto de Ciencias Nucleares, Universidad Nacional Aut\'onoma de M\'exico, A.P. 70-543, M\'exico D.F. 04510, M\'exico.}

\author{Marcelo Salgado}
\email{marcelo@nucleares.unam.mx}
\affiliation{Instituto de Ciencias Nucleares, Universidad Nacional Aut\'onoma de M\'exico, A.P. 70-543, M\'exico D.F. 04510, M\'exico.}%


\date{\today}


\begin{abstract}
Using a 3+1 decomposition of spacetime, we derive a new formula to compute the gravitational light shifts as measured 
by two observers which are normal to the spacelike hypersurfaces defining the foliation. This formula is quite general 
and is also independent of the existence of Killing fields. Known examples are considered 
to illustrate the usefulness of the formula. In particular, we focus on the Sachs-Wolfe effect that arises in a 
perturbed Friedman-Robertson-Walker cosmology.
\end{abstract}
\maketitle

\section{Introduction}
\bigskip
\noindent 
In a flat spacetime the frequency shifts of light (FSL) or Doppler effect, arises due to the relative motion between the source and 
the detector (see Ref.~\cite{rotshift} for {\it rotational} FSL as opposed to translational FSL), while in 
a curved spacetime, the FSL can be due to several factors. The more striking one is perhaps the FSL associated with 
the time dilation due to the presence of a (strong) gravitational field in the neighborhood of a clock. 
Clearly if a source and a detector are moving relative to each other in a curved spacetime, kinematical and gravitational 
FLS can be combined into a non trivial fashion. This is precisely what happens in the detection of light coming from many astrophysical 
sources. If the source is far away from us, then the FLS can have cosmological contributions in addition to the kinematical, 
thermal and local gravity contributions. The flat rotation curves of (spiral) galaxies \cite{FRC} is a notable example of the kind of 
results that can be obtained by measuring the FSL. Spectroscopy is usually the basic tool to measure the FSL.

One of the most important effects that allows one to test the predictions of general relativity (or any other metric theory of gravity) are 
the gravitational FSL (redshift or blue shift)~\cite{PoundRebka}. In particular, these shifts can be easily computed when the spacetime 
possesses several symmetries. The fact that the projection of a Killing vector field along a geodesic tangent vector is constant along that 
geodesic, simplifies considerably the calculation of such shifts. When {\it a priori} Killing fields are absent or when the spacetime has 
only approximate symmetries (as in the case of a perturbed symmetric spacetime), this calculation might not be straightforward. The so called 
Sachs-Wolfe effect~\cite{SachsWolfe} that arises in a perturbed Friedman-Robertson-Walker (FRW) cosmology and that we discuss below 
(Sec. \ref{sec:sachs-wolfe}) represents 
a prominent example of such a case.

The observation of light shifts provide some information about the geometry or strength of the gravitational field hosting the light source 
and might also help for distinguishing between different gravity theories in the strong field regime~\cite{DeDeo}. 
In the case of cosmology, the Sachs-Wolfe effect can help to distinguish between 
alternative metric theories since for instance the potentials $\psi$ and $\phi$ used in the Newtonian gauge 
for scalar perturbations~\cite{Mukhanov,Bardeen} might differ (not only in sign) at the time of decoupling, but also in magnitude 
depending on the gravitational theory one is dealing with~\cite{Zhang}. For example, under the standard linear perturbation theory 
in pure general relativity, when one discards the 
anisotropic stresses of matter at the time of decoupling one has $|\psi|\approx |\phi|$, and the Sachs-Wolfe effect then 
depends only on one of these potentials (cf. Eq.~[\ref{SW}] ). However, in alternative theories of gravity like 
scalar-tensor theories and modified $f(R)$ theories one has 
{\em a priori} $|\psi|\neq |\phi|$~\cite{Mukhanov,Zhang} since the effective energy-momentum tensor associated with those theories 
is not necessarily isotropic ({\it i.e.} it might include non-diagonal terms at the time of decoupling). 
So the Sachs-Wolfe effect can be a useful tool to validate, bound or even rule out alternative theories.

The Sachs-Wolfe effect has two contributions: one which is produced 
at the last scattering surface (the primary or primordial Sachs-Wolfe effect) and another one which is due to the 
effective FSL produced when the photons encounter local gravitational potentials that evolve in time during their trip towards 
our detectors (the integrated Sachs-Wolfe or ISW). Both effects are reported in the most recent 
Cosmic-Background-Radiation (CBR) observations and appear at large angular scales in the sky ($\gtrsim 10^o$). In particular, the primordial 
Sachs-Wolfe effect (the quadrupole contributions to the CBR) has to be contrasted with the contributions 
due to the relative motion of our galaxy with respect to the CBR frame 
(the dipole contributions) or due to the plasma oscillations (FSL noticeable at small angular scales). 
The primordial effect is almost scale invariant and so it is often referred to as the Sachs-Wolfe {\it plateau} 
in the CBR power spectrum. 

In this report we obtain a novel formula for computing the FSL 
associated with a certain family of observers. In the context of the 3+1 decomposition of 
spacetime, such observers are hypersurface orthogonal and are usually called Eulerian. 
In this context, the interpretation of the origin of FSL will depend on the identification or not
of the Eulerian observers with the observers associated with the underlying symmetries of a spacetime. 
For instance, in a flat spacetime the formula leads to zero FSL if the observers are {\it inertial} (see below). 
However, if one chooses non inertial observers for which the metric components are not {\it canonical}, then the formula can 
lead to non trivial FSL. In this sense, the formula presented below does not really distinguish between FSL coming from
observers in a real gravitational field ({\em i.e.} observers whose world lines live in a curved spacetime) 
and FSL coming from accelerated observers in a flat spacetime. Here we see one clear illustration of the Einstein's equivalence 
principle. Two specific examples of this situation are to be found in the Rindler spacetime (a portion of Minkowski spacetime as viewed by 
accelerated observers whose hypersurfaces of simultaneity have Euclidean topology)
and in the Milne universe (a portion of Minkowski spacetime as viewed by expanding observers whose hypersurfaces of simultaneity have 
hyperbolic topology). One should then be careful in the interpretation of FSL outcome and not to confuse results which are 
coordinate dependent with results that are observer dependent (see Refs.~\cite{Faraoni,Groen2007} for a discussion on this 
issue and also for the interpretation of cosmological FSL versus Doppler effect and the Milne example as well).

The formula does not rely on the existence of Killing vector fields nor on a specific 
metric theory of gravitation. However, when the former are present, the 
formula naturally reproduce the well known results (see Sec~\ref{sec:examples}). On the other hand, when Killing fields are absent, 
we show the usefulness of the formula by using as model example the case of a linearly perturbed FRW spacetime 
which leads to the Sachs-Wolfe effect.

\section{The formula }
\label{sec:2}
\bigskip
We shall assume that a spacetime $(M,g_{ab})$\footnote{Here we use the Wald's convention~\cite{Wald} where 
Latin indices $a,b,c$ refer to abstract indices, while Greek indices refer to components and run $0-3$. We use 
Latin indices ($i,j=1,2,3$) to denote spatial components.}, is globally hyperbolic and thus that it admits a foliation by a family of 
spacelike Cauchy hypersurfaces $\Sigma_t$, parametrized by a global time function, $t$. 
The normal $n^a$ to these hypersurfaces is a future pointing time-like vector field 
({\em i.e.} $n_a n^a =-1$) which defines a family of observers called {\it Eulerian} 
(see Refs.~\cite{Wald,3+1} for a thorough introduction to the 3+1 formalism).\\

On the spacetime manifold $M$, one is given a local coordinate system $x^\mu$ such that $x^\mu= (t,x^i)$ is a local coordinate system 
adapted to the foliation of $M$ where $x^i$ is a local {\it spatial} coordinate system of the embedded manifold $\Sigma_t$. 
Therefore, with respect to these coordinates $n^a= (1/N) (\partial/\partial t)^a + (N^i/N) (\partial/\partial x^i)^a$ or 
simply $n^\mu= (1/N,N^i/N)$ and $n_a= -N\nabla_a t$ ({\em i.e.} $n_\mu= (-N,0,0,0)$ ), where $N$ is the {\it lapse} function and $N^a$ is the 
{\it shift vector}. This latter is orthogonal to $n^a$ and has only spatial contravariant components. In terms of the 3+1 decomposition the 
metric reads
\begin{equation}
 ds^2= -\left(N^2- N_i N^j\right)dt^2 - 2N_i dx^i dt + h_{ij} dx^i dx^j\,,
\end{equation}
where $h_{ij}$ stands for the 3-metric components (see below).

Now, consider two Eulerian observers (one associated with the source of light and the other one associated with the detector) 
located on $\Sigma_t$ so that their point location $p_e$ and $p_d$ have spacetime coordinates $(t,x_e^i)$ and $(t, x_d^i)$. 
A light signal is emitted from $p_e$ which is detected at $p_d$ at time $t+\Delta t$. The photon's null geodesic four vector $k^a$ is such that 
$\frac{dx^\mu}{d\lambda}=k^\mu$, where $x^\mu(\lambda)=(t(\lambda), x^i (\lambda))$ provides the path of the photon in terms of the local 
coordinate system and $\lambda$ is an affine parameter. Now, the light frequency measured by any of the 
Eulerian observers at some point $p$ is given by $\omega=-k^an_a |_p$. So $\omega (x^\mu (\lambda) )$ is a scalar field that changes 
{\it smoothly} along the photon's path. We consider then that at point $p_e$ the photon's path has coordinates given 
by $x^\mu (\lambda)$ while at the detection point $p_d$ has coordinates $x^\mu (\lambda+\Delta \lambda)$. 
In other words, we consider that along the photon's path 
there is always an Eulerian observer measuring its frequency (at the intersection point between the photon null geodesic and the observer 
path). That is, a family of Eulerian observers can be represented locally by a congruence of lines or orbits 
that intersects the photon's path at different points. The frequency shift between the emitted and detected light signal is then given by
\begin{equation}
 \omega_d - \omega_e = \omega (x^\mu (\lambda +\Delta \lambda) )- \omega (x^\mu (\lambda )) \,\,\,.
\end{equation}
The ``instantaneous'' frequency shift along the photon's path is then given by
\begin{equation}
 {\rm lim}_{\Delta\lambda\rightarrow 0} \frac{\omega (x^\mu (\lambda +\Delta \lambda) )- \omega (x^\mu (\lambda))}{\Delta \lambda}
 = \frac{d\omega}{d\lambda}= \frac{dx^\mu}{d\lambda}\nabla_\mu \omega \,\,\,.
\end{equation}
This {\it local} FSL is associated with the light frequency measured by Eulerian observers which are {\it infinitesimally} closed to each other. 
From the above definition we have 
\begin{equation}
\label{deromega}
\frac{d\omega}{d\lambda}= k^a\nabla_a\omega =-k^a\nabla_a (k^bn_b) = -k^ak^b\nabla_a n_b \,\,\,.
\end{equation}
where in the last step we used the fact that photons follow geodesics $k^a\nabla_a k^b=0$. On the other hand, the term 
$\nabla_a n_b$ is related to the {\it extrinsic curvature} $K_{ab}$ (or 
 second fundamental form) of $\Sigma_t$  by 
 $\nabla_a n_b= -K_{ab}- n_aa_b$ ~\cite{3+1}, where $a_a = n^c \nabla_c n_a$ is the 
 four acceleration of the normal observers. We remind the reader that the extrinsic curvature is given by 
 \footnote{It is to note that in Wald's book~\cite{Wald}, the extrinsic curvature is defined with the opposite sign.}
 $K_{ab}:= -\frac{1}{2} \: {\cal L}_{\mathbf{n}} h_{ab}$, where ${\cal L}_{\mathbf{n}} $ stands for the Lie derivative along $n^a$, and 
 $h_{ab}= g_{ab} + n_a n_b$ is the {\it induced} metric (or 3-metric) on $\Sigma_t$ ~\cite{Wald,3+1}. In fact it is not difficult to 
prove that the four acceleration $a_a$  can be written in terms of the {\it lapse} function $N$ as follows $a_a= h_{a}^{\,\,b}\nabla_b\ln N 
 = D_a \ln N$, where $D_a$ is the covariant derivative compatible with the 3-metric ~\cite{Wald,3+1}. 
 By using these results in Eq.~(\ref{deromega}) we obtain
 \begin{equation}
\label{deromega2}
\dfrac{d\omega}{d\lambda}=  k^a k^b K_{ab}- \omega k^a D_a\ln N \,\,\,.
\end{equation}
Furthermore, the projector 
$h^a_{\,\,b} = \delta^a_{\,\,b} + n^a n_b $ can be used to define a 3-vector $^3 k^a :=h^a_{\,\,b}\;k^b$ 
which is tangent to $\Sigma_t$ and orthogonal to $n^a$. Moreover we can define 
$^3\hat k^a=\frac{^3 k^a}{\omega}$  which is normalized ({\em i.e.} $^3\hat k^a \,^3\hat k^b h_{ab} = 1$) since 
$k^a$ is null. The vector $k^a$ can then be decomposed as follows 
$k^a= \,^3 k^a + \omega n^a=\omega \left(^3\hat k^a + n^a\right)$. Inserting this expression for $k^a$ 
in Eq.~(\ref{deromega2}) and using the fact that $n^a K_{ab}\equiv 0\equiv n^a D_a \ln N $  we obtain 
\begin{eqnarray}
\label{formula0}
\dfrac{d\omega}{d\lambda}&=& \;^3\,\!k^a\;^3\,\!k^b K_{ab}-\omega\;^3\,\!k^a D_a\ln N \nonumber\\
&=&\omega^2\left(\;^3\hat k^a\;^3\hat k^b K_{ab}-\;^3\hat k^a D_a\ln N\right) \,\,\,.
\end{eqnarray}
An alternative way of writing Eqs.~(\ref{deromega2}) and (\ref{formula0}) is as follows
\begin{equation}
\label{formula}
\frac{1}{\omega}\dfrac{d\omega}{d\lambda}= \omega\left(\;^3\hat k^a\;^3\hat k^b K_{ab} + 
{\cal L}_{\mathbf{n}}\ln N \right) -\frac{d \ln N}{d\lambda} \,\,\,.
\end{equation}
where we used $D_a\ln N=\nabla_a \ln N + n_an^b \nabla_b \ln N$ and then 
$k^a D_a\ln N=d\ln N/d\lambda - \omega {\cal L}_{\mathbf{n}}\ln N$.

In this way Eq.~(\ref{formula}) provides the most general formula for the local gravitational light shift. 
In order to obtain a finite frequency shift ({\em i.e.} a frequency shift as measured by two Eulerian observers separated by a finite spatial distance) 
one needs to integrate the previous equation along the photon's path. Now, a physical interpretation to the previous equation can be given, but before 
we proceed, it will be useful to write the following explicit expression for the spatial components of the extrinsic curvature in terms of the 3-metric 
and the {\it shift vector} $N^a$~\cite{3+1}
\begin{equation}
\label{Kij}
 K_{ij}= -\frac{1}{2N} \left(\partial_t h_{ij}  +D_i N_j + D_j N_i \right) \,\,\,.
\end{equation}
We are now in position to give some insight about the different terms entering in Eq.~(\ref{formula}). 
In {\it static} situations where one usually identifies the function $t$ defining the spacelike hypersurfaces $\Sigma_t$ 
with the parameter associated with the hypersurface orthogonal Killing vector field $\xi^a = (\partial/\partial t)^a$, it turns 
$n^a= \xi^a/N$ and the shift vector is identically null. In this case the term within parenthesis in Eq.~(\ref{formula}) 
also vanishes identically [cf. Eq.~(\ref{Kij}) ]. Therefore when the spacetime is static the Eulerian observers are naturally identified 
with {\it static observers} and the FSL measured by them reduces to the relationship $\omega_e/\omega_d= N(x^i_d)/N(x^i_e)$,
where $x^i_e,x^i_d$ refer to the spatial coordinates of the points where the light was emitted and detected respectively. 
As a result of this, the lapse function is often referred to as {\it the redshift factor}. Note that in general $N\neq 1$, 
which implies that the Eulerian observers are not necessarily geodesic ({\em i.e.} the four acceleration $a_a= D_a \ln N$ is not null in general). 
In the next section we use the FRW cosmology as an example where the natural Eulerian observers are geodesic, 
in which case, $a_a$ vanishes identically. On the other hand, the term within parenthesis in Eq.~(\ref{formula}) provides 
additional frequency shifts when the spacetime changes in time (cf. Eq.~[\ref{Kij}] ) or when it is 
stationary. Indeed, in a stationary situation the timelike Killing field is not hypersurface orthogonal but the metric is time independent. 
Therefore, in this situation and when one identifies the Eulerian observers' coordinates with those associated with the symmetries of a stationary 
spacetime, it turns that the formula Eq.~(\ref{formula}) accounts for two contributions to the FSL: one is due to the fact that the Eulerian observers 
are not geodesic and therefore are accelerating [this contribution is again due to the last term at the r.h.s of Eq.~(\ref{formula})]; the second 
contribution arises from the terms which involve the shift vector $N^a$ [cf. Eq.~(\ref{Kij})]. This additional contribution to the FSL is due to 
the so called {\it dragging of inertial frames}. In the next section we give a specific example that illustrates these interpretations. 
Another simple but non trivial situation arises in FRW cosmology where $N=1$ and $N^a\equiv 0$. In this case the FSL is only due to time variations 
of the gravitational field ({\em i.e.} due to the expansion of the Universe; see next section). Finally, we shall consider the 
Sachs-Wolfe effect (see Sec.~\ref{sec:sachs-wolfe}) where the time and spatial variations of the gravitational field combine to give a FSL.

One last comment is in order. This is in fact related to a potential confusion between coordinate and observer dependent effects 
(see Ref.~\cite{Faraoni} for a further discussion on this issue). 
One could in fact be surprised in giving a physical interpretation to the terms appearing in 
Eq.~(\ref{formula}), since after all, one thing that one learns in the analysis of the 3+1 decomposition of spacetime is that 
both $N$ and $N_i$ define the coordinate gauge and therefore that their meaning is not physical. However, 
precisely different choices of $N$ and $N_i$ define the kind of observers we use to ``coordinetize'' the spacetime and, in the present context, 
the kind of observers we use to compute FLS as well. As mentioned before, the Rindler spacetime~\cite{MTW} is a simple example of this situation. In Rindler 
coordinates the lapse function is given by  $N= (1+ g x)$ (where $g$ is the proper acceleration of the observer whose worldline is associated 
with the coordinate $x=0$), while the rest of the metric components are trivial. Therefore one finds a non zero FSL due to the term with 
$d\ln N/d\lambda$ in Eq.~(\ref{formula}). This FSL is due to the fact that two Rindler observers accelerate in a different way depending on 
their relative positions, and not due to the curvature of spacetime\footnote{In this case the FSL formula leads directly to 
$\omega_e/\omega_d = (1+ gx_d)/(1+ gx_e)$; when $x_e\rightarrow -1/g$, an infinite shift arises because of the Rindler horizon. Something similar 
occurs in the Schwarzschild spacetime when one of the static observers is located arbitrarily near the event horizon; see Sec.~\ref{sec:examples} 
[cf. Eq.~(\ref{Sch})].}. 
Actually the Rindler spacetime represents only a portion of Minkowski spacetime. Note however that in ordinary Minkowski Cartesian coordinates, 
$N=1$, $N_i\equiv 0$ and $h_{ij}= \delta_{ij}$. Then the FSL is exactly zero. This means that the Eulerian observers in question are {\it inertial}. 
That is, they are in relative rest and so there are no FSL whatsoever. Moreover, if one consider a boosted family of observers, these also are inertial and 
the metric components for them takes exactly the same form as the metric for the other family of observers. Then for this second family of 
boosted observers there are no FSL either. So the formula given by Eq.~(\ref{formula}) is by construction unable to account for the FSL 
({\em i.e.} Doppler shifts) due to a relative motion of (local) inertial observers. 

\section{Examples}
\label{sec:examples}

\bigskip

\noindent{\bf Static and spherically symmetric spacetime}: Consider the line element $ds^2= -N^2(r)dt^2+A^2(r)dr^2+r^2d\Omega^2$. 
In this case the normal observers to $\Sigma_t$ are static, and so $n^a=(1/N)\xi ^a$; where $\xi^a= 
(\partial/\partial t)^a$ is the timelike hypersurface orthogonal Killing field. 
Moreover, for this metric $K_{ij}\equiv 0\equiv {\cal L}_{\mathbf{n}}\ln N$. From Eq.~(\ref{formula}) one easily finds:
 \begin{equation}
 \frac{\omega_e}{\omega_d}=\frac{N(r_d)}{N(r_e)}\,\,\,.
 \end{equation}
In the case of the Schwarzschild solution we recover the usual expression\footnote{However, in alternative theories of gravity 
(notably, in scalar-tensor theories) the lapse function may not have 
an analytic expression when the scalar field outside a compact object is not trivial~\cite{DeDeo,Salgado}.} 
 \begin{equation}
\label{Sch}
  \frac{\omega_e}{\omega_d} = \sqrt{\frac{1-\frac{2M}{r_d}}{1-\frac{2M}{r_e}}} \,\,\,.
 \end{equation}
 {\bf FRW spacetime}: The line element is given by 
  \begin{equation}
  \label{FRW}
  ds^2=-dt^2+a^2(t)\left[\frac{dr^2}{1-kr^2}+r^2d\Omega^2\right]\,\,\,.
  \end{equation}
 In this case the normal observers to $\Sigma_t$ are comoving, and so $n^a= (\partial/\partial t)^a$ ({\em i.e.} $N\equiv 1$). 
The 3-metric $h_{ij}$ can be read off from Eq.~(\ref{FRW}), and from Eq.~(\ref{Kij}) we find 
$K_{ij}=-\frac{\dot a}{a}h_{ij}$ and ${\cal L}_{\mathbf{n}}\ln N\equiv 0\equiv d \ln N/d\lambda $. 
 Replacing these expressions in Eq.~(\ref{formula}) and using $d/d\lambda= \omega d/dt$ one easily finds the familiar expression
  \begin{equation}
   \frac{\omega_e}{\omega_d}=\frac{a(t_d)}{a(t_e)}\,\,\,.
  \end{equation}
Depending on the cosmological model ({\em e.g.} matter content) and the theory at hand, one has explicit solutions for $a(t)$
\footnote{It is interesting to mention that for $a(t)= t$ and $k=-1$, the metric (\ref{FRW}) corresponds actually to a
flat spacetime (without matter)~\cite{Groen2007,Peacock}. The Eulerian observers are 
expanding in a non trivial fashion and the spacelike hypersurfaces $\Sigma_t$ have hyperbolic topology. 
This spacetime is referred to as the {\it Milne Universe}~\cite{Milne} and the comoving coordinates cover only a portion of 
Minkowski spacetime. In fact the hypersurfaces $\Sigma_t$ are not really Cauchy surfaces 
of the whole Minkowski spacetime but only of that portion covered by the comoving coordinates. Therefore it is only that portion 
that can be foliated by the hypersurfaces $\Sigma_t$.}.

{\bf Stationary and axisymmetric spacetime}: 
This case is perhaps a little more interesting than the previous ones, because the metric induces dragging effects on the light shifts. 
Let us then consider the following line element:
\begin{equation}
ds^2  = -\left(N^2-N_\varphi N^\varphi \right) dt^2 -2N_\varphi dtd\varphi + h_{ij} dx^i dx^j\,.
\end{equation}
where all the metric components are time and $\varphi$ independent but depend on the two coordinates 
$(x^1,x^2)$ which can be chosen to be of spherical or cylindrical type. 
Apart from the fact that $h_{\varphi i} \equiv 0$ (for $i=1,2$), the explicit form of the 3-metric $h_{ij}$ 
does not concern us since it will not be necessary in the calculation of the FSL. 
In this example, the Eulerian observers have four velocity given by 
$n^a = (1/N)(\partial/\partial t)^a + (N^\varphi/N) (\partial/\partial \varphi)^a$, where in fact, 
$\xi^a= (\partial/\partial t)^a$ and $\psi^a= (\partial/\partial \varphi)^a$ are the timelike and the 
spacelike Killing fields, respectively, which are associated with the time and axial symmetries. 

A straightforward calculation leads to
\begin{equation}
^3 k^a\;^3 k^b K_{ab}= -\frac{\;^3 k_\varphi \,k^\mu}{N}\partial_\mu N^\varphi= 
-\frac{\;^3 k_\varphi}{N} \frac{dN^\varphi}{d\lambda}\,\,\,.
\end{equation}
where $^3k_a:= h_{ab}\,^3\,\!k^b=g_{ab}\,^3 k^b$ and in the first equality we used 
$^3 k^i= k^i$ (for $i\neq \varphi$) since $N^\varphi$ is the only non-null component of the 
shift vector. In fact since $k^\mu \partial_\mu N^\varphi \equiv 0$ (for $\mu=t,\varphi$)
by the stationary and axisymmetry conditions, those terms do not contribute to $k^\mu\partial_\mu N^\phi$, 
but we have retained them in order to explicitly obtain the last equality. Moreover, since 
$^3k_a= k_a -\omega n_a= g_{ab} k^b -\omega n_a$, and using 
$n_i \equiv 0$ (for $i=1,2,3$) 
then $^3k_\varphi= g_{\varphi b} k^b\equiv k_\varphi$. The quantity $L:= g_{ab} \psi^a k^b= g_{\varphi b} k^b$ 
which is conserved along the photon's path is identified with the 
photon's {\it angular momentum}. In this way Eq.~(\ref{formula}) leads to 
the following differential equation 
\begin{equation} 
\frac{1}{\omega}\dfrac{d\omega}{d \lambda}=  -\frac{L}{\omega N}\frac{dN^\varphi}{d\lambda}  
- \frac{d\ln N}{d\lambda} \,\,\,.
\end{equation}
This equation can be easily integrated and when evaluated at the emission and detection points we find
\begin{equation}
\label{formulaxi}
\frac{\omega_e}{\omega_d}=\frac{N_d}{N_e}\frac{\left[1-\frac{L}{E}N^\varphi_e\right]}
{\left[1-\frac{L}{E}N^\varphi_d\right]} \,\,\,,
\end{equation}
where $E= -k^a \xi_a$ is an integration constant which is identified with the photon's {\it energy}~\footnote{If the spacetime 
is asymptotically flat $E$ is the photon's energy as measured by an Eulerian
observer at spatial infinity.} and which is also conserved along the photon's path. 
Here the subscripts $e$ and $d$ at the r.h.s mean that the quantities have to be computed at the points of emission and detection 
with coordinates $(x^1_e,x^2_e)$ and $(x^1_d,x^2_d)$ for any $t,\varphi$. 
Of course this calculation can be done in a few steps using the Killing vector fields from the 
start\footnote{We have $\omega= -k_a n^a$ where $n^a = \xi^a/N + N^\varphi \psi^a/N$ then 
$\omega= (E/N)\left(1 - \frac{L}{E} N^\varphi\right)$ where $E= -k_a \xi^a$ and $L= k_a \psi^a$ are 
constants along the photon's path. Evaluating $\omega$ at the points of emission and detection we recover 
Eq.~(\ref{formulaxi}). It is interesting to mention that for this kind of spacetimes the Eulerian observers 
are also called ZAMOs (Zero Angular Momentum Observers)~\cite{Bardeen73} 
since $L_Z:= n^a\psi_a\equiv 0$, as one can check.}.

\section{Sachs-Wolfe effect}
\label{sec:sachs-wolfe}
\bigskip
Let us considered {\it scalar linear perturbations} of the FRW metric in the 
Newtonian gauge~\cite{Mukhanov}
\begin{equation}
\label{new}
ds^2= -(1+2\phi) dt^2 + (1+2\psi)\cero h_{ij}dx^i dx^j\,\,\,,
\end{equation}
where $\cero h_{ij}$ corresponds to the unperturbed FRW 3-metric of Eq.~(\ref{FRW}) and $|\phi|,|\psi|\ll1$. 
Up to first order, the lapse function is $N= 1 + \phi$, and the 3-metric $h_{ij}= (1+2\psi)\cero h_{ij}$. 
So from Eq.~(\ref{Kij}) the perturbed extrinsic curvature reads 
$K_{ij}=-\dfrac{\dot a}{a}\cero  h_{ij}+ \left(-\partial_t\psi+\phi\dfrac{\dot a}{a}-2\psi\dfrac{\dot a}{a}\right)\cero h_{ij}$, 
where the first term provides the zero-order contribution as we saw in the second example. Moreover, we can write 
$^3\hat k^a= \,\!^3\hat k^a_0 + \,\!^3\hat k^a_1$ where the lower indices $0,1$ refer to zero and first order respectively, 
so that the normalization condition is verified at zero order and perturbatively as well: 
$^3\hat k^a \,^3\hat k^b h_{ab}=1 = \,\!^3\hat k^a_0 \,^3\hat k^b_0 \cero h_{ab}$. In this way, 
up to first order, $^3\hat k^a \,^3\hat k^b \cero h_{ab}= 1- 2\psi$. These preliminary results allow us to find 
$^3\hat k^i \,^3\hat k^j K_{ij}= -\dfrac{\dot a}{a} + \phi \dfrac{\dot a}{a} -\partial_t \psi$. On the other hand, 
$d{\rm ln}N/d\lambda= d\phi/d\lambda$ and ${\cal L}_{\mathbf{n}}\ln N= n^\mu\nabla_\mu{\rm ln}N= 
\partial_t \phi$. Collecting all these partial results into Eq.~(\ref{formula}) we obtain:
\begin{equation}
\frac{1}{\omega}\frac{d\omega}{d\lambda}= \omega\left(-\frac{\dot a}{a} - 
\frac{\partial\psi}{\partial t}+ \frac{\dot a}{a}\phi 
+ \frac{\partial\phi}{\partial t}  \right) -\frac{d\phi}{d\lambda} \,\,\,.
\end{equation}
Now, since $\omega= -n_a k^a= N k^t$, then up to first order $k^t/\omega= 1/N= 1- \phi$, therefore, 
$\frac{d\omega}{d\lambda}= \frac{d\omega}{dt} k^t= \frac{d\omega}{dt} \omega (1-\phi)$, and similarly 
$\frac{d\phi}{d\lambda}= \frac{d\phi}{dt}\omega (1-\phi)$. In this way one obtains the 
following expression valid up to first order,
\begin{equation}
\label{SW0}
\frac{d\omega}{\omega dt}=-\frac{\dot a}{a}-\left(\frac{\partial\psi}{\partial t}
-\frac{\partial\phi}{\partial t}\right) -\frac{d\phi}{dt}\,\,\,.
\end{equation}
Note the cancellation of the term $\frac{\dot a}{a}\phi$. The first term at the r.h.s of Eq.~(\ref{SW0}) is 
associated with the unperturbed FSL. Finally, we can write the perturbed 
frequency as $\omega=\omega_0+\omega_1$, and so $\frac{d\omega}{\omega dt} =
\frac{d{\rm ln} \omega_0}{dt} + \frac{d{\rm ln}(1+\omega_1/\omega_0)}{dt}$. Now, for the unperturbed frequency 
we have $\frac{d{\rm ln} \omega_0}{dt}= - \frac{d{\rm ln} a}{dt}$. On the other hand, up to first order 
$\frac{d{\rm ln}(1+\omega_1/\omega_0)}{dt}= \frac{d}{dt}(\frac{\omega_1}{\omega_0})$. In this way we have proved
that up to first order $\frac{d\omega}{\omega dt}+ \frac{\dot a}{a}= \frac{d}{dt}(\frac{\omega_1}{\omega_0})$.

Eq.~(\ref{SW0}) can now be integrated with respect to $t$ to obtain the usual Sachs-Wolfe effect expression:
\begin{equation}
\label{SW}
\left.\frac{\delta T}{T_0}\right|^{t_d}_{t_e}=\phi(\vec{x_e},t_e)-\phi(\vec{x_d},t_d)+ \int^{t_d}_{t_e}
\frac{\partial D(\vec{x}(t),t)}{\partial t} dt
\end{equation}
where we defined $\frac{\delta T}{T_0}= \frac{\omega_1}{\omega_0}$, 
as the relative temperature perturbations ($T_0\sim \omega_0$ is the unperturbed temperature of the FRW Universe, 
and $\delta T\sim \omega_1$ is the temperature perturbation) and $D(\vec{x},t):= \phi(\vec{x},t)-\psi(\vec{x},t)$. 
The last integral has to be evaluated along the photon's path
\footnote{One can take into account not only scalar perturbations but also first-order vector and tensor perturbations as well. 
In order to do so a convenient and simple gauge which generalizes the Newtonian gauge is the Poisson 
gauge~\cite{Bertschinger,Bruni}. In such a gauge the perturbed metric reads:
\begin{equation}
\label{pois}
ds^2= -(1+2\phi) dt^2 -2N_i^T dx^idt+ \left[ (1+2\psi)\cero h_{ij}+ 2{\cal H}_{ij}^{TT} \right]dx^i dx^j\,\,\,,
\end{equation}
where $N_i^T$ means that the shift perturbation is transverse (${\cero D}^i N_i^T=0$, where ${\cero D}^i$ stands for 
the 3-covariant derivative compatible with to the non-perturbed metric $\cero h_{ij}$) and ${\cal H}_{ij}^{TT}$ is transverse and 
traceless (${\cero D}^i {\cal H}_{ij}^{TT}=0= \cero h^{ij}{\cal H}_{ij}^{TT}$). Notice that in this case we have six physical 
perturbations (two scalars $\phi$ and $\psi$, two associated with the transverse vector $N_i^T$, and two which provide the two polarization modes 
of the gravitational waves associated with the transeverse traceless tensor ${\cal H}_{ij}^{TT}$). 
The contribution of the vector and tensor perturbations to the Sachs-Wolfe effect is straightforward. In this case and up to first order, 
$K_{ij}= K_{ij}^S - \partial_t {\cal H}_{ij}^{TT} - {{\cero D}_{(i}} N_{\,j)}\,\!\!\!\!\!\!^T$ , where $K_{ij}^S$ is the 
extrinsic curvature up to first order which includes only the scalar perturbations as in the main 
text. So using Eq.~(\ref{formula}) and the results of the main text one obtains
\begin{equation}
\label{SWP}
\left.\frac{\delta T}{T_0}\right|^{t_d}_{t_e}= \left.\left(\frac{\delta T}{T_0}\right)^S\right|^{t_d}_{t_e}
 - \int^{t_d}_{t_e}\, ^3\hat k^i_0\,^3\hat k^j_0\left[\frac{\partial {\cal H}_{ij}^{TT} 
(\vec{x}(t),t)}{\partial t} + \cero D_{i} N_{j}^T (\vec{x}(t),t)\right] dt\,\,\,,
\end{equation}
where the first term at the r.h.s is given by the r.h.s of Eq.(\ref{SW})
(see Ref.~\cite{Magueijo,Hwang1999} for an alternative derivation which includes scalar, vector and tensor perturbations using a 
gauge-invariant and a general-gauge formalisms, respectively). 
}.

The primordial temperature fluctuations associated with the potential $\phi$ give rise to the ordinary Sachs-Wolfe effect, which corresponds to 
the redshift of light due to the ``climbing'' of photons through the potential $\phi(\vec{x_e},t_e)$ at the last scattering 
surface. The term $\phi(\vec{x_d},t_d)$ does not really contributes to the anisotropies since it is associated with the local gravitational 
field around the detector which contributes isotropically to the temperature perturbations. The last term which involves the integral 
is associated with the temperature perturbations due to the time variations of the 
potentials along the photon's path. It is called the {\it integrated Sachs-Wolfe} (ISW) effect. Notice that in Eq. (\ref{SW}) 
temperature fluctuations due to peculiar velocities (Doppler shifts) are absent due to the limitations of the formula (\ref{formula}), 
as we have stressed before.

In general relativity and in absence of anisotropic stresses, the Einstein equations imply $\psi=-\phi$, and then 
$D(\vec{x},t)= 2\phi(\vec{x},t)$ which leads to the usual expression for the  Sachs-Wolfe effect.
In fact since the l.h.s of Eq.~(\ref{SW}) is $(\delta T/T_0)_{t_d}-  (\delta T/T_0)_{t_e}$ one can show $ (\delta T/T_0)_{t_e}\approx -2\phi/3$ 
(for adiabatic perturbations in a matter dominated epoch, and $k=0$ universe)~\cite{Mukhanov,Hwang1999,White,Hwang2002} and so 
$(\delta T/T_0)_{t_d}= \phi(\vec{x_e},t_e)/3 + {\rm ISW}$. The actual primordial temperature anisotropies measured today between photons coming from two 
different points (angles) at the last scattering surface is $\frac{\Delta T(\vec{x_e^1},\vec{x_e^2})}{T}:= \frac{1}{3}\left[\phi(\vec{x_e^1},t_e)- 
\phi(\vec{x_e^2},t_e) \right] + {\rm ISW}^1_2 $.

As mentioned in the introduction, in alternative metric theories of gravity the effective energy-momentum tensor associated with these theories 
is not {\it a priori} isotropic ({\it i.e.} it is spatially non-diagonal) at the last scattering surface, and therefore  $|\psi|\neq |\phi|$. 
So, this can have observational consequences in the CBR angular power spectrum of temperature anisotropies~\cite{Zhang,Zhao}.

\section{Discussion}

\bigskip
Based on a 3+1 decomposition of spacetime, 
we have presented a novel formula to compute the frequency shifts of light between two observers which are in general non geodesic. 
The formula does not account for the Doppler (kinematical) effects which arises when one of the observers is geodesic and the other is not or when 
the two observers are connected by Lorentz transformations (notably in flat spacetime). That is, when one considers two observers with 
different four velocities $n^a$ and $u^a$ such that {\it locally} $n_au^a= -\Gamma$, where $\Gamma$ is the Lorentz factor relative to both observers. 
In this sense, the formula only accounts for the ``gravitational'' FSL arising when the spacetime is genuinely curved or in certain cases where 
one uses non inertial coordinates to describe the Minkowski spacetime (accelerated observers). The formula does not rely on the existence of 
Killing vector fields, however, when the latter are present, the formula allows one to recover the well known results. Furthermore, it is theory 
independent since it uses only the kinematical properties of the spacetime but not the field equations.

Since historically gravitational light shifts have been computed in several fashions and using different 
techniques, it is important to comment about some of the methods used in the past and to contrast them with our formula. For instance, Wald~\cite{Wald} 
proposes as an exercise a derivation similar to ours but only for the case of FRW cosmology. Also in the cosmological context, 
one can find different deductions~\cite{Faraoni,Groen2007} where the interpretation of Doppler or gravitational shifts are 
discussed. For stationary spacetimes, Gr\o n~\cite{Groen1980} obtained a formula which also takes into account the Doppler 
effects due to the relative motion of observers ({\it e.g.} one of whom is geodesic and the other is not). 
Using Killing vector fields FLS formulae are obtained in Ref.~\cite{Harvey}. Ellis~\cite{Ellis}, 
obtained a FSL formula using a 1+3 congruence formalism (as opposed to the 3+1 employed here) 
(see also Ref.~\cite{Ellis98} for a further review). In all such treatments, the 
Sachs-Wolfe effect is not computed whether because the authors focus only on FRW cosmology or because their formalism applies only to stationary 
situations. On the other hand, Dunsby~\cite{Dunsby} based on the 1+3 covariant approach~\cite{Ellis89} does analyze the Sachs-Wolfe effect for 
scalar perturbations (see also Refs.~\cite{Russ,Panek}).

In our case, we showed that our formula leads in a more straightforward and geometrical fashion to the Sachs-Wolfe effect. 
Thanks to its generality and simplicity we consider that the formula can be useful in several situations of physical interest.

\begin{acknowledgements}
We warmly thank R. Sussman for enlightening discussions. This work was supported in part by CONACyT grant 
CB-2007-01-082787, and by DGAPA-UNAM grants IN119309-3 and IN115310. 
\end{acknowledgements}






\end{document}